\documentclass[10pt,aps,prd,nofootinbib,superscriptaddress,twocolumn]{revtex4}
\usepackage{amssymb,amsmath,latexsym,graphics, graphicx,epsfig} 
\usepackage{latexsym}
\usepackage{amssymb}
\usepackage{epsfig,amsmath,graphics}
\usepackage{multirow}
\usepackage{appendix}
\usepackage{comment}

\newcommand{\gsim}{\lower.7ex\hbox{$\;\stackrel{\textstyle>}{\sim}\;$}}
\newcommand{\lsim}{\lower.7ex\hbox{$\;\stackrel{\textstyle<}{\sim}\;$}}
\def\LL{{\cal L}}

\def\OO{{\cal O}}

\newcommand{\TeV}{\,\mathrm{TeV}}
\newcommand{\GeV}{\,\mathrm{GeV}}
\newcommand{\MeV}{\,\mathrm{MeV}}
\newcommand{\keV}{\,\mathrm{keV}}

\newcommand{\half}{{\frac{1}{2}  }}

\newcommand{\hc}{\text{ h.c. }}

\newcommand{\Tr}{{\text{ Tr }}}

\newcommand{\Br}{{\text{ Br}}}
\newcommand{\eff}{{\text{eff}}}

\newcommand{\bef}{\begin{figure}[htbp]\begin{center}}
\newcommand{\eef}{\end{center}\end{figure}}

\newcommand{\td}{{\text{d}}}
\newcommand{\pid}{{\pi_\td}}
\newcommand{\munu}{{\mu\nu}}
\newcommand{\rhod}{{\rho_\td}}
\newcommand{\thpv}{{ \theta_{\text{P}\hspace{-0.085in}\not\hspace{0.07in}} }}
\newcommand{\pv}{{\text{P}\hspace{-0.085in}\not\hspace{0.07in}} }
\newcommand{\Ld}{{\Lambda_\td }}

\newcommand{\D}{\mbox{$D\hspace{-0.15in}\not\hspace{0.12in}$}}

\begin{document}

\pagestyle{plain}

\title{
\begin{flushright}
\mbox{\normalsize SLAC-PUB-13841}
\end{flushright}
\vskip 15 pt

Parity Violation in Composite Inelastic Dark Matter Models}

\author{Mariangela Lisanti and Jay G. Wacker }

\affiliation{
Theory Group, SLAC,  Menlo Park, CA 94025
}

\begin{abstract}
Recent experimental results indicate that the dark matter sector may have a non-minimal structure with a spectrum of states and interactions.  Inelastic scattering has received particular attention in light of DAMA's annual modulation signal.  Composite inelastic dark matter (CiDM) provides a dynamical origin for the mass splittings in inelastic dark matter models.  We show that higher dimensional operators in the CiDM Lagrangian lead to an admixture of inelastic and elastic scattering in the presence of parity violation.  This scenario is consistent with direct detection experiments, even when parity violation is nearly maximal.  We present an effective field theory description of such models and discuss the constraints from direct detection experiments.  
The CiDM model with parity violation has non-trivial phenomenology because of the multiple scattering channels that are allowed.
  \end{abstract}
 \maketitle

\section{Introduction}

Recent direct and indirect searches for dark matter hint that the dark sector may have non-minimal structure and interactions.  This is in sharp contrast with  the standard scenario of weakly interacting massive particles in which the dark matter is the lightest neutral state in a spectrum and interacts elastically off of Standard Model (SM) particles.  The results of the DAMA experiment provide an example of an anomaly that challenges the standard dark matter picture \cite{Bernabei:2005hj,Bernabei:2008yi}.  In particular, if DAMA's measured annual modulation arises from inelastic dark matter (iDM), then DAMA can be reconciled with all other null results from direct detection experiments \cite{TuckerSmith:2001hy,Chang:2008gd}.  The presence of multiple states that lead to inelastic interactions may indicate novel dynamics in the dark sector.  For example, iDM requires an $\OO(100\keV)$ splitting between the dark matter states, which may be a sign that dark matter is composite \cite{Alves:2009nf, Kaplan:2009de,Nussinov:1985xr,Chivukula:1989qb,Bagnasco:1993st,Khlopov:2008ki, Kouvaris:2008hc,Ryttov:2008xe,Mohapatra:2001sx,Kribs:2009fy}.  

Composite inelastic dark matter (CiDM) is a recent proposal that provides a dynamical origin for the 100 keV mass splitting \cite{Alves:2009nf}.  CiDM models have a ground state degeneracy that is split by the hyperfine interaction.
In \cite{Alves:2009nf}, a minimal CiDM model was proposed where a new strong gauge group confines at low energies and the quarks that
are charged under the new strong gauge group form ``dark hadrons'' after confinement. 
The interactions between the Standard Model and the dark sector are mediated by a kinetically-mixed $U(1)_\td$.  

The parity of the $U(1)_\td$ current determines how visible the dark matter is to
the Standard Model.  The minimal CiDM model conserved parity and the axial-vector current interaction  led to only inelastic interactions.  However, parity can be explicitly broken by the dynamics of the  new strong gauge group through a $\Theta$ term.  
Parity violation leads to elastic interactions that arise from charge-radius scattering and are phenomenologically different from typical elastic dark matter interactions because the charge-radius scattering is suppressed at low nuclear recoil energies.   The recoil spectrum of this model looks strikingly similar to standard iDM models due to the suppression from low recoil energy interactions.  CiDM models provide a new framework for studying kinematic scenarios where several types of scattering events are allowed.  

This article presents the theory of CiDM models with parity violation.  Sec.~\ref{Sec: Models} describes the low energy effective theory  in terms of ``dark mesons."  Sec. \ref{Sec: Direct Detection} computes the direct detection phenomenology using a global fit while marginalizing over the uncertainty in the dark matter velocity distribution function.  Sec.~\ref{Sec: DarkPhoton} discusses constraints arising from QED tests and summarizes prospects for collider searches.  

\section{Models of CiDM}
\label{Sec: Models}

In this section, the effective field theory for CiDM models
is reviewed and the consequences of parity violation in the new strong sector is explored. 
The high energy theory is a two flavor  $SU(N_c)$ gauge theory with a Lagrangian given by
\begin{eqnarray}
\LL&=& \LL_{\text{SM}} + \LL_{\text{CiDM}}  \\ \nonumber
\LL_{\text{CiDM}} &=& -\frac{1}{2} \Tr G_{\td\, \mu\nu}^2+  \bar{\Psi}_L i \D \Psi_L  + \bar{\Psi}_H i \D \Psi_H\\ \nonumber
&&+ m_L \bar{\Psi}_L \Psi_L + m_H \bar{\Psi}_H \Psi_H 
\end{eqnarray}
where $\Psi_a$, $a= L, H$, are Dirac fermions that are fundamentals under the strong gauge sector and $G_{\td\,\mu\nu}$ is the $SU(N_c)$ gauge field strength.
In Atomic Inelastic Dark Matter (AiDM) \cite{Kaplan:2009de}, the strong gauge sector is replaced by an Abelian gauge
group where  $\Psi_a$ are charge $\pm 1$, respectively.   If $\Psi_a$ have
chiral gauge charges, then the $m_a$ arise through a symmetry breaking interaction (\emph{i.e.}, $m_a = y_a \langle \phi \rangle$).
If the theory is asymptotically free as in \cite{Alves:2009nf}, then the theory will confine at a scale $\Lambda_\td$.
The bound states will be approximately Coulombic if $m_L \gsim \Lambda_\td$ and the resulting spectrum is qualitatively similar to AiDM.

To have the appropriate relic density, the dark matter must either be very heavy ($\gsim 30\TeV$) or the dark matter density must be generated non-thermally, possibly linked to baryogensis.   $\OO(30\TeV)$ inelastic dark matter is not compatible with CDMS's null results \cite{Akerib:2005za, Akerib:2005kh,Ahmed:2008eu} if it fits the DAMA signal and therefore the relic abundance of the dark matter needs to be generated non-thermally.  Assuming that there is a cosmological asymmetry generated early in the Universe 
between heavy quarks and light anti-quarks,
\begin{eqnarray}
n_{H} - n_{\bar{H}} = -n_{L} + n_{\bar{L}} \neq 0,
\end{eqnarray}
the dominant component of the dark matter will be in dark mesons with a single heavy quark \cite{Alves:2009nf}. 

The dark matter is the ground state of a $\Psi_{\text{H}} \bar{\Psi}_{\text{L}}$ meson and will be denoted as
$\pid$.  The $\pid$ is a spin 0, complex scalar with parity $-1$.   Dark matter scattering is primarily
a transition to the complex, spin 1 meson, $\rhod$.  The parity of this state is
\begin{eqnarray}
P \rhod_\mu = (-1)^\mu \rho_\mu \qquad (-1)^\mu = \begin{cases} \;\;1& \mu=0\\ -1& \mu = 1, 2, 3\end{cases} .
\end{eqnarray}
The mass splitting between the $\pid$ and the $\rhod$ arises through the hyperfine interaction
and is suppressed when $m_H \gg m_L, \Lambda_\td$.  In particular,
\begin{eqnarray}
\label{Eq: Mass Splitting}
\nonumber
&& \frac{\kappa \Ld^2}{m_H}, \qquad m_L \ll \Ld\\
 \delta m=m_\rhod - m_\pid \simeq& \\
\nonumber
&&\frac{ \lambda_\td^4m_L^2}{N_c m_H}\qquad m_L\gg \Ld,
\end{eqnarray}
where $\lambda_\td =  N_cg_{\td}^2/4\pi$ is the 't Hooft coupling with $N_c=1$ applying to Abelian gauge groups and $\kappa$ is an $\OO(1/N_c)$ constant.  Note that $m_{\pid} \sim m_{\text{H}}$ in the heavy quark mass limit.  For mass splittings $\OO(100$ keV) and a dark matter mass near the weak scale, the confinement scale is $\sim 100$ MeV.

The dark matter mesons interact through a massive spin 1 gauge
field $A_\td^\mu$ that kinetically mixes with $U(1)_Y$ \cite{Holdom, Arkani-Hamed:2008,Katz:2009qq,Morrissey:2009ur,Goodsell:2009xc,Cheung:2009qd}
\begin{eqnarray}
\nonumber
\LL_{\text{Gauge}}&=& -\frac{1}{4}  F^2_\td  +\frac{ \epsilon}{2} F_\td^{\mu \nu}B_{\mu\nu} \\
\LL_{\text{Higgs}} &=& |D_\mu\phi|^2 - \lambda( |\phi|^2 -\half f_\phi^2)^2,
\end{eqnarray}
where $D^\mu\phi = \partial^\mu \phi - 2 ig_\td A_\td^\mu\phi$. 
After $\phi$ acquires a vev, $f_\phi$, $A_\td$ becomes massive and the mixing
between the dark sector and the Standard Model can be diagonalized.
The $\epsilon$ mixing between the hypercharge field strength and $U(1)_\td$ is the source of the interactions between the Standard Model fermions and the dark matter.
Assuming that 
\begin{eqnarray}
m_{A_\td}= 2 g_\td f_\phi \ll m_{Z^0}
\end{eqnarray}
after electroweak symmetry breaking, the couplings can be diagonalized.
The interactions relevant for dark matter scattering
are given by
\begin{eqnarray}
\mathcal L_{\text{Int}} = \left( J_{\text{d}}^{\mu}+ \epsilon c_{\theta} J_{\text{EM}}^{\mu}\right)
{A_\td}_{\mu}   ,\label{eq:SMinteractions}
\end{eqnarray}
where   $c_\theta= \cos\theta_{\text{w}}$, $J_{\text{EM}}$ is the electromagnetic current and $J_{\text{d}}$  is the current of the dark quarks. 
A more complete analysis of the interactions is given in Sec. \ref{Sec: DarkPhoton}.
 Anomaly cancellation restricts the charge assignments of the two dark quarks leaving only three anomaly-free possibilities for the current of the dark quark sector.

\subsection{Axially Charged Quarks}

The types of interactions that are allowed depend on whether $J_\td$ is an axial or vector current.
In \cite{Alves:2009nf} and \cite{Kaplan:2009de}, $J_\td$ is an axial vector current.
The only anomaly-free axial charge assignment in terms of Weyl spinors is
\begin{eqnarray}
\begin{array}{|c|cccc|}
\hline
\text{Axial}&\psi_H&
\psi_H^c & 
\psi_L &
\psi_L^c \\
\hline
SU(N_c)&  \square& \overline{\square}& \square & \overline{\square}\\
q_{U(1)_\td}& +1 &+1 &-1&-1\\
\hline
\end{array},
\end{eqnarray}
where the Dirac spinors $\Psi_a = ( \psi_a, \bar{\psi}_a^c)$.
The masses of the quarks arise from  $U(1)_\td$ breaking through the Higgs mechanism
\begin{eqnarray}
\LL_{\text{Yuk}} =  y_H \phi \psi_H \psi_H^c + y_L \phi^\dagger \psi_L \psi_L^c +\hc .
\end{eqnarray}
Because both the mass of the quarks and the $A_\td$ arise from the vev of $\phi$, there is a hierarchy between the gauge couplings and the Yukawa couplings
\begin{eqnarray}
\frac{m_{A_\td}}{m_{\pid}} = \frac{ 2 g_\td}{ y_H} .
\label{eq: mA}
\end{eqnarray} 
Fitting DAMA requires $m_{\pid} \sim 100$ GeV, while $m_{A_{\td}}$ can in principle take on a range of values from
$10 \MeV$ to $100\GeV$.  Eq.~\ref{eq: mA} implies that the gauge coupling for the axial sector may be small in comparison to the Standard Model gauge couplings because $y_H$ is capped by perturbativity at $\OO(1)$.

The effective operators describing the interactions of the $\pid - \rhod$ system are
\begin{eqnarray}
\nonumber
\LL_{\text{Axial }\eff} &=&   d^{\text{a}}_{\text{in}} m_\pid \pid^\dagger \rho^\mu_\td A_{\td \mu} \\
\nonumber
&&+\frac{c^{\text{a}}_{\text{in}}}{\Lambda_\td}  \pid^\dagger \partial^\mu \rho^\nu_\td F_{\td\,\mu\nu}\\
\nonumber
&&+ \frac{d^{\text{a}}_{\text{el}}}{\Lambda^2_\td}( \pid^\dagger \partial_\mu \pid + \rho^\nu_\td{}^\dagger \partial_\mu \rhod_\nu) \partial_\nu \tilde{F}^{\mu\nu}_\td\\
&& + c^{\text{a}}_{\text{el}} \rhod^\dagger_\mu \rhod_\nu \tilde{F}^{\mu\nu}_\td 
+\hc.
\label{Eq: Leff Axial}
\end{eqnarray}
The operators with coefficients denoted by $d$ are suppressed by a factor of the relative velocity $v_{\text{rel}}$, while the 
ones denoted by $c$ are not velocity suppressed.  The elastic scattering operator for the $\pid$ 
is dimension 6 and velocity suppressed, resulting in an overall suppression of the elastic to inelastic scattering rate of $v_{\text{rel}}^2$.

\subsection{Vectorially Charged Quarks}

There are two anomaly-free charge assignments for vectorially charged dark matter:
one gives the composite dark matter a charge
and the other leaves it neutral.  Charged dark matter
will have an enormous scattering rate and will look qualitatively similar to
standard elastic dark matter.  The charge assignment that leaves the dark
matter neutral will only scatter off higher moments of the charge distribution
and will be suppressed at low recoil energy.  The charge assignments for the neutral dark matter theory are
\begin{eqnarray}
\begin{array}{|c|cccc|}
\hline
\text{Neutral Vector}&\psi_H&
\psi_H^c & 
\psi_L &
\psi_L^c \\
\hline
SU(N_c)&  \square& \overline{\square}& \square & \overline{\square}\\
q_{U(1)_\td}& +1 &-1 &+1&-1\\
\hline
\end{array} .
\end{eqnarray}
With these charge assignments,  $A_\td$ couples to a vector current and the allowed operators are
\begin{eqnarray}
\nonumber
\LL_{\text{Vector }\eff} &=&\frac{d^{\text{v}}_{\text{in}}}{\Lambda_\td}  \pid^\dagger \partial^\mu \rho^\nu_\td \tilde{F}_{\td\,\mu\nu}\\
\nonumber
&&+ \frac{c^{\text{v}}_{\text{el}}}{\Lambda^2_\td} ( \pid^\dagger \partial_\mu \pid + \rho^\nu_\td{}^\dagger \partial_\mu \rhod_\nu)\partial_\nu F^{\mu\nu}_\td\\
&& + d^{\text{v}}_{\text{el}} \rhod^\dagger_\mu \rhod_\nu F^{\mu\nu}_\td  +\hc .
\label{Eq: Leff Vector}
\end{eqnarray}
$d$ denotes operators that are velocity suppressed and $c$ denotes unsuppressed
operators.  The leading operator that is not velocity suppressed is the elastic charge-radius operator,
but this is a dimension 6 operator.    
Recent work on form factor-suppressed inelastic transitions indicates that this type of scattering
may be an explanation for DAMA \cite{Chang:2009yt,Feldstein:2009tr}.  The primary difference
between form factor elastic scattering dark matter and iDM
is the existence of a threshold in iDM.  The next section illustrates that it is possible
to have dark matter dominantly scatter inelastically and have a residual form factor elastic contribution.

\subsection{Parity Violation}
\label{Sec: PV}

In the two models above, parity determined the interactions of the dark meson fields; however,
parity is not a fundamental symmetry of nature.
If parity is broken, both charge-radius scattering and inelastic scattering are allowed without a velocity suppression.
This is quite natural in strongly coupled CiDM models because 
  CP violation arises from the dynamics of the strong sector through the term
\begin{equation}
\LL_{\pv} = \Theta_\td\!\! \Tr G_\td \tilde{G}_\td,
\end{equation}
and results in mixing between states of different parity.   The size of $\Theta_\td$ is not necessarily related
to the size of $\Theta_{\text{QCD}}$ and in principle $\Theta_\td$ could be $\OO(1)$.  

Because the $\Theta_\td$ term is a total derivative, its effects only appear non-perturbatively.  The dominant effect of the CP violation is to cause a small mixing between states
of different parity.  In QCD, for example, the $\pi^0$ with $I^G (J^{P})=1^- (0^{-})$ and the $a_0$ with $I^G(J^{P})=1^-(0^{+})$ mix in the presence of a $\Theta_{\text{QCD}}$ term.  A similar process will happen in the dark sector.  When $m_L\lesssim \Ld$, the mixing angle between fields of opposite parity is given by 
\begin{eqnarray}
\sin \thpv \sim \Theta_\td \frac{m_L}{\Ld} .
\end{eqnarray}
The mixing vanishes in the limit where $m_L \rightarrow 0$ because the $\Theta_{\td}$ term can be removed by a chiral rotation of the $\Psi_{\text{L}}$.  If $m_L \gg \Ld$, the mesons form Coulombic bound states and the  mixing angle is given by
\begin{eqnarray}
\sin \thpv =  
\frac{\langle \pid| H_{\pv} |a_{0\,\text{d}}\rangle}{m_{a_{0\,\text{d}}}-m_{\pid}}
\simeq \frac{\Theta_\td \Ld}{\lambda_\td^2 m_L},
\end{eqnarray}
where the matrix elements of the perturbing CP-violating Hamiltonian is set by the non-perturbative scale where the effects of $\Theta_\td$ are not exponentially suppressed.  As $m_L\rightarrow \infty$, the CP-violating effects decouple and parity violation vanishes.  Therefore, even if $\Theta_\td \sim \OO(1)$, its effects on the interactions of the dark mesons might be small if $m_{\text{L}} \rightarrow 0, \infty$.  Maximal parity violation occurs when  $m_L\simeq \Ld$.    

With an axially coupled $U(1)_{\td}$, the $\pid - a_{0\,\text{d}}$ interaction becomes an elastic charge-radius operator
with parity violation:
\begin{eqnarray}
\nonumber
\LL _{\pid a_{0\,\text{d}}}\! = \! \frac{c^{a}_{\text{el}}}{\Lambda^2_\td}\pi_{\td}^\dagger \partial_\mu a_{0\,\text{d}} \partial_\nu F^{\mu\nu}_\td 
\!\rightarrow  \!
\frac{c^{a}_{\text{el}}}{2\Lambda^2_\td} \sin 2 \thpv \,\pi_\td^\dagger \partial_\mu\pid \partial_\nu F^{\mu\nu}_\td .
\end{eqnarray}
Therefore, the effects of $\Theta_\td$ can be estimated by replacing the field strengths  in Eq. \ref{Eq: Leff Axial} and Eq. \ref{Eq: Leff Vector} with\footnote{This only applies to the field strengths, $F^\munu_\td$, not the gauge potentials, $A^\mu_\td$, whose interactions are constrained by gauge invariance.} 
\begin{eqnarray}
F^\munu_\td \rightarrow  \cos 2\thpv F^\munu_\td + \sin 2\thpv \tilde{F}^\munu_\td.
\end{eqnarray}
Therefore, turning on parity violation in the strong sector allows admixtures of vector and axial vector interactions.  The ratio of elastic to inelastic cross sections becomes a free parameter.
The next section will show that an upper bound of $\thpv \lsim 0.08$ is necessary to avoid direct detection constraints assuming that all $c, d\sim \OO(1)$ and $m_N\sim m_\pid$.

\section{Direct Detection Phenomenology}
\label{Sec: Direct Detection}

Novel features in the direct detection phenomenology of composite models arise because the dark matter has a finite size $\Lambda^{-1}_{\text{d}} \gg m_\pid^{-1}$.
The cross section is suppressed by an effective form factor  
when a  neutral bound state interacts with momentum $|\vec{q}\,| \ll \Lambda_\td$ \cite{Pospelov:2000bq}.
States with nonzero spin have multipole interactions with the field.  These moments vanish for states with zero spin; scalar states that can only couple through the charge-radius and polarizability interactions are the dominant scattering mechanisms.  For the dark pion scattering off the SM, the charge-radius interaction dominates over the polarizability interaction, which is suppressed by an additional factor of the mixing parameter $\epsilon^2$.  

The charge-radius is the effective size of the $\pid$ probed by the dark photon.  In the limit of small momentum transfer $|\vec{q}\,| \ll \Lambda_{\td}$, the wavelength of the dark photon is too long to probe the charged constituents of the composite state and the scattering rate is suppressed.  Elastic charge-radius scattering cannot be the sole contributor to the direct detection signal due to constraints from current null experiments.  However see  \cite{Chang:2009yt,Feldstein:2009tr} for examples on how form factors can reconcile DAMA with the null experiments. 

The dominant scattering is inelastic and there is a subdominant elastic component that accounts for a fraction of the total scattering rate.
Specifically, the differential scattering cross section is  
\begin{eqnarray}
\frac{d\sigma}{dE_R} &=& \left(  \thpv^2 \frac{4 m_N^2  E_R^2 \kappa }{(m_\pid \delta m)^2}  + \frac{m_N  E_R }{2 m_\pid \delta m}\right)
\frac{d\sigma_0}{d E_R}, \nonumber
\label{Eq: Sigma}
\end{eqnarray}
where  $m_N$ is the mass of a nucleus with charge $Z$ recoiling with energy $E_R$, $ c_{\text{in}}^a, c_{\text{el}}^a=1$ of Sec. \ref{Sec: PV} and
\begin{eqnarray}
\frac{d \sigma_0}{d E_R} =  \frac{ 8 Z^2\alpha m_N}{  v^2} \frac{1}{f_{\eff}^4}
\frac{|F_{\text{Helm}}(E_R)|^2}{\left(1+ 2m_N E_R/m_{A_\td}^2\right)^2} .
\label{eq: xsection}
\end{eqnarray}
The scattering operators couple the dark matter states coherently to the nuclear charge, and the Helm form factor accounts for loss of the coherence at large recoil
\begin{eqnarray}
|F_{\text{Helm}} (E_R)|^2 = \left(\frac{3j_1(|q|r_0)}{|q|r_0}\right)^2 e^{-s^2|q|^2},
\end{eqnarray}
where $s=1$ fm, $r_0=\sqrt{r^2-5s^2}$, and $r=1.2A^{1/3}$ fm  \cite{Helm:1956zz}.

\begin{figure}[b] 
   \centering
   \includegraphics[width=3.4in]{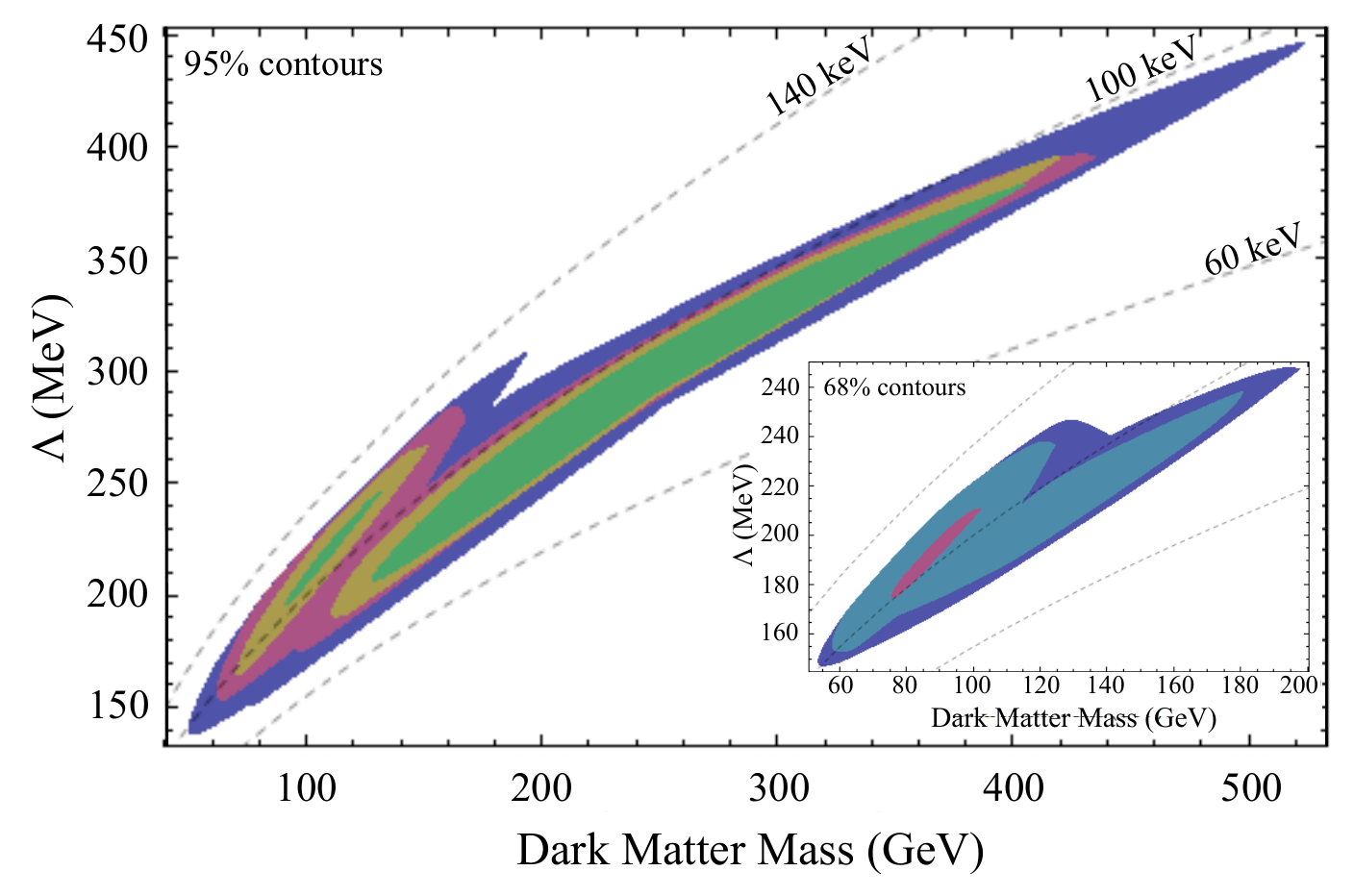} 
   \caption{95\% contours in $m_{\pid} - \Lambda_\td$ parameter space for $\thpv = 0.00, 0.06, 0.07, 0.08$.  For this figure, $\kappa = 1/4$ and $m_{A_\td} = 1$ GeV.  The dashed lines show contours of $\delta m$ in keV. The inset shows the 68\% confidence regions for $\thpv = 0.00, 0.04, 0.06$ for the same $\kappa$ and $m_{A_\td}$.   The colors correspond to $\thpv=$ $0.00$ (blue), $0.04$ (teal), $0.06$ (magenta), $0.07$ (yellow), $0.08$ (green).  }
   \label{fig: masslambda}
\end{figure}

The differential cross section depends on the confinement scale $\Lambda_{\td} = \sqrt{m_\pid \delta m/ \kappa} $, the mass of the dark photon $m_{A_{\td}}$, and the couplings of the effective theory
\begin{equation}
f^2_\eff = \frac{m_{A_\td}^2}{ \kappa g_d \epsilon}, 
\label{eq: feff}
\end{equation}
where $\kappa$ is the $\OO(1/N_c)$ constant defining the mass difference from Eq. \ref{Eq: Mass Splitting}.  

To determine the preferred region of parameter space for CiDM models, a global $\chi^2$ analysis was performed that included the results from all current direction detection experiments.  This procedure is outlined in \cite{upcoming, Lisanti:2009vy} and is summarized here.  The differential scattering rate per unit detector mass is 
\begin{equation}
\frac{dR}{dE_R} = \frac{\rho_0}{m_{\pid} m_N}\! \int\!\! d^3 v\; f(\vec{v} + \vec{v}_e)\, v \frac{d\sigma}{d E_R},
\end{equation}
where $\rho_0= 0.3$ GeV/cm$^3$ is the local dark matter density and $\vec{v}_e$ is the velocity of the Earth in the galactic rest frame.  There are significant uncertainties in the dark matter velocity distribution function $f(v)$, and constraints on the particle physics model can vary wildly depending on the particular choice of benchmark halo model.  To find the full scope of allowed CiDM models, we marginalize over a parameterized velocity distribution function of the form:
\begin{eqnarray}
f(v) \propto \exp\left( \frac{v}{v_0}\right)^{2\alpha}\!\! -\; \exp\left( \frac{v_{\text{esc}}}{v_0}\right)^{2\alpha},
\end{eqnarray}
where the parameters are constrained to be within
\begin{eqnarray}
\nonumber
&200 \text{ km/s} \le v_0 \le 300 \text{ km/s}&\\
\nonumber
&500 \text{ km/s} \le v_{\text{esc}} \le 600 \text{ km/s}&\\
&0.8 \le \alpha \le 1.25\, . &
\end{eqnarray}
These values are motivated by observational constraints \cite{Lewin:1995rx,Smith:2006ym} and analytic approximations to the Via Lactea results \cite{Diemand:2006ik, Fairbairn:2008gz,MarchRussell:2008dy}.   

The global $\chi^2$ fit is performed by marginalizing over the six unknown parameters of the dark matter and halo model: $m_{\pid}, \delta m, f_{\text{eff}}, v_0, v_{\text{esc}},$ and $\alpha$.  The measurements used in the $\chi^2$ fit are
 the first twelve bins of DAMA's modulation amplitude, as well as a single high energy bin from $8\text{ keVee}$ to $12 \text{ keVee}$ \cite{Bernabei:2005hj,Bernabei:2008yi,Chang:2008xa}.  In addition to DAMA's signal, the dark matter predictions are required to not supersaturate any observation from null experiments at the 95\% confidence level.  The null experiments included in the analysis are: CDMS \cite{Akerib:2005za, Akerib:2005kh,Ahmed:2008eu}, ZEPLINII \cite{Alner:2007ja}, ZEPLINIII \cite{ZEPLINIII}, CRESSTII \cite{Angloher:2004tr,CRESSTII}, and the new XENON10 inelastic dark matter analysis \cite{XENON10, XENON10iDM}.  

\begin{figure}[tb] 
   \centering
   \includegraphics[width=3.4in]{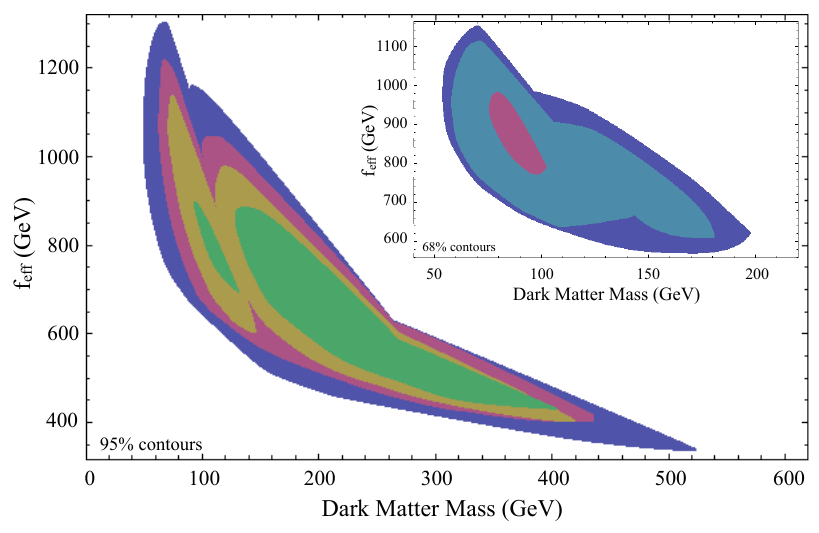} 
   \caption{95\% confidence limit regions of $m_\pid - f_\eff$ for $\thpv = 0.00, 0.06, 0.07, 0.08 $.  For this figure, $\kappa = 0.25$ and $m_{A_\td} = 1 \GeV$.  The inset shows the 68\% confidence regions for $\thpv = 0.00, 0.04, 0.06$ for the same $\kappa$ and $m_{A_\td}$.}
   \label{fig: massfeff}
\end{figure}

The $1\sigma$ and $2\sigma$ allowed regions in the $m_{\pid}-\Lambda_{\td}$ and $m_{\pid} - f_{\text{eff}}$ spaces are shown in Fig.~\ref{fig: masslambda} and~\ref{fig: massfeff}, respectively.  The minimal $\chi^2$ has a value of $4.61$ and the corresponding point is listed as model CiDM$_1$ in Table \ref{Tab: Benchmarks}.  The $1\,\sigma$ and $2\,\sigma$ regions are set by 
\begin{eqnarray}
\chi^2 < \chi^2_{\text{min}} + \Delta \chi^2,
\end{eqnarray}
where $\Delta \chi^2 (1\sigma) = 7.0$ and $\Delta \chi^2 (2\sigma) = 12.6$.  Therefore, the 68\% and 95\% regions are set by requiring that $\chi^2 \le 11.6, 17.2$, respectively.  As the fraction of form factor elastic scattering increases relative to the inelastic contribution, the allowed regions in Fig.~\ref{fig: masslambda} and~\ref{fig: massfeff} each separate into two.  At 95\% confidence, dark matter masses with $m_{\pid} \gtrsim 200$ GeV correspond to ``slow'' velocity distribution functions where $v_0 \lsim 225 \text{km/s}$ and $ \alpha \gsim 1.15$.  One benchmark model is shown in Table \ref{Tab: Benchmarks} as CiDM$_3$.  However, the correlations between the dark matter mass and the velocity distribution parameters are far weaker in the low mass region ($m_\pid \lesssim 200$ GeV).  

The mass of the dark photon is related to the mixing parameter $\epsilon$ as 
\begin{equation}
\epsilon = \frac{m_{A_{\td}}^2}{g_d f^2_{\text{eff}}} = \frac{2 \sqrt{2} m_{A_\td} m_{\pid}}{y_H f_{\text{eff}}^2}.
\end{equation}
For theories with dominant inelastic scattering, $f_{\text{eff}} \sim \OO(700 \text{ GeV})$ and $m_{\pid} \sim \OO(100 \text{ GeV})$ to satisfy both the DAMA and null experiments.  Therefore, keeping $y_H \simeq 1$
\begin{equation}
\epsilon = \OO(10^{-4}) \frac{m_{A_{\td}}}{1 \text{ GeV}},
\end{equation}
which corresponds well with the results of the $\chi^2$ global fit.  Fig.~\ref{Fig: Dark Photon Limits} shows the $1\sigma$ and $2\sigma$ regions in the $m_{A_{\td}} - \epsilon$ parameter space allowed by all current direct detection experiments.  A benchmark value of $y_H =1$ is chosen; the contours shift to larger $\epsilon$ for smaller Yukawa coupling.  

Light $A_\td$ alter the fit to the DAMA spectrum because the propagator suppresses high momentum scattering events.  The momentum transfer needed to explain the highest energy bin with a statistically signifiant annual modulation rate in the DAMA  spectrum ($E_R = 5$ keVee) is
\begin{eqnarray}
\label{Eq: Momentum Transfer}
|\vec{q}\,| = \sqrt{ \frac{2 m_{I} E_R}{q_I}} \simeq 120 \MeV.
\end{eqnarray}
If the mass of the dark photon is less than 120 MeV, its propagator in (\ref{Eq: Sigma}) suppresses the scattering rate in the high energy bins.
The suppression of the high momentum transfer events can be compensated if the mass splitting, $\delta m$, grows larger; however this 
 eventually forces $f_{\text{eff}}^{-1}$ to become large, increasing the allowed values of $\epsilon$.  These effects are shown in Fig. \ref{Fig: Dark Photon Limits}.  A low $m_{A_\td}$ benchmark model is shown as CiDM$_4$ in Table. \ref{Tab: Benchmarks}.
The following section presents constraints on the allowed parameter region arising from indirect and direct searches for the dark photon.

\begin{table}
\begin{tabular}{|c||c|c|c|c|}
\hline
& CiDM$_1$& CiDM$_2$& CiDM$_3$ &CiDM$_4$\\
\hline\hline
$m_\pid$& 72 GeV& 75 GeV & 234 GeV & 162 GeV\\
$\delta m$& 109 keV& 105 keV&91 keV & 126 keV\\  
$\Lambda_\td$&177 MeV& 177 MeV& 292 MeV& 286 MeV\\
$f_\eff$& 738 GeV& 846 GeV &563 GeV& 268 GeV \\
$\epsilon$&$3.7\times 10^{-4}$&$3.0\times 10^{-4}$&$2.1\times 10^{-3}$&$3.8\times 10^{-4}$ \\
$m_{A_\td}$& 1 GeV& 1 GeV& 1 GeV& 60 MeV\\
$\thpv$&0.00& 0.04& 0.04& 0.06\\
\hline
$v_0$& 272 km/s& 273 km/s & 202 km/s& 280 km/s\\
$v_{\text{esc}}$&510 km/s& 501 km/s& 558 km/s& 501 km/s\\
$\alpha$&0.86& 0.82 &1.30& 0.98\\
\hline\hline
$\chi^2$& 4.6&6.2 &12.8&9.9 \\
\hline
\end{tabular}
\caption{\label{Tab: Benchmarks}  Four benchmark models showing different regions of parameter space. CiDM$_1$ corresponds to the best-fit point.  CiDM$_2$ shows a representative mixture of inelastic and subdominant elastic scattering.  CiDM$_3$ shows the larger mass window with slow halo parameters.  CiDM$_4$ shows a light $m_{A_\td}$ model.}
\end{table}

 \section{Searches for the Dark Photon}
\label{Sec: DarkPhoton}

\begin{figure}[b] 
\centering
\includegraphics[width=3.4in]{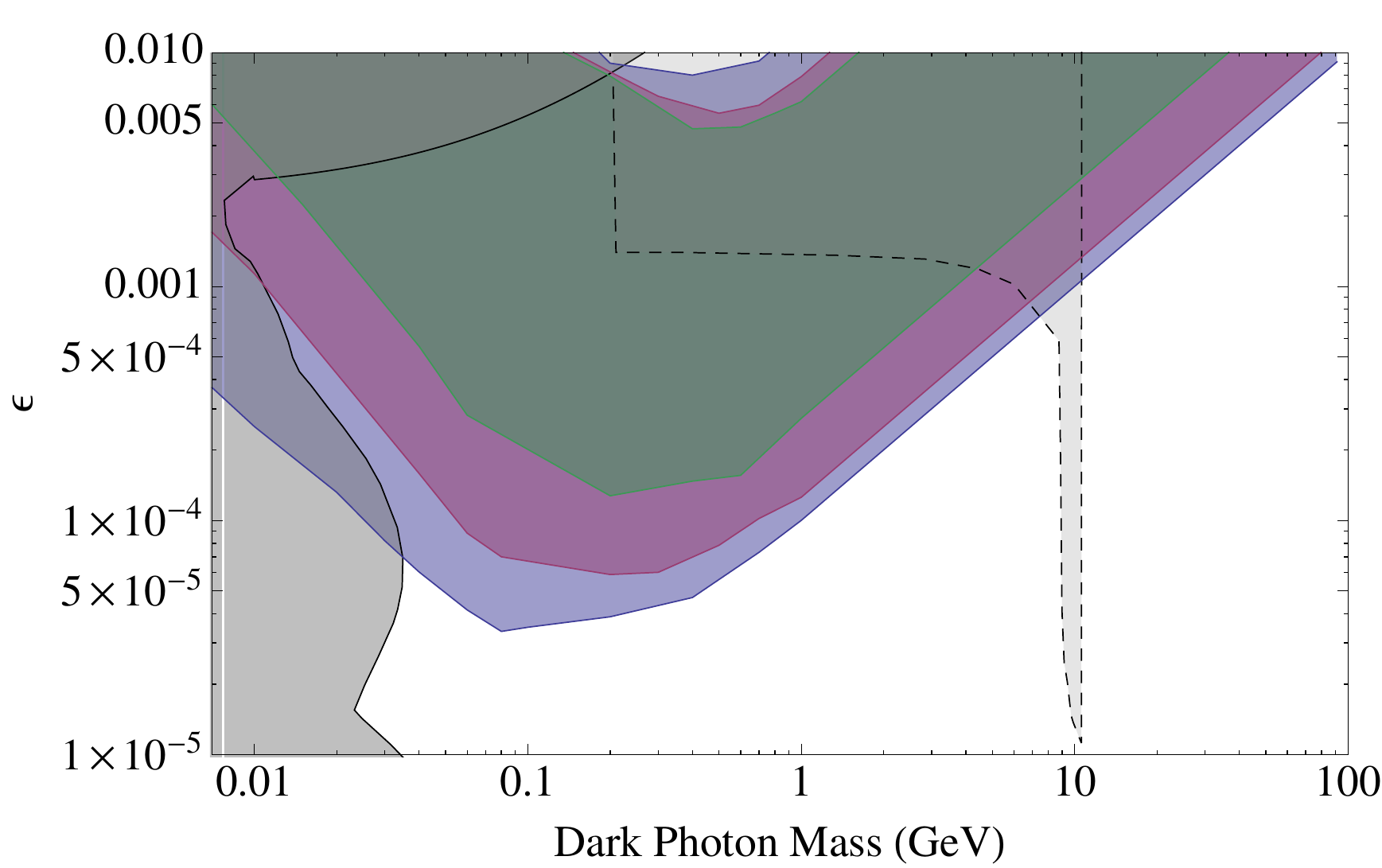} 
\caption{The 95\% limits on $m_{A_\td} - \epsilon$ with $y_H=1$ for  $\thpv = 0.00$ (blue), $0.06$ (magenta), $0.08$ (green).  The dark gray regions are excluded by limits on the $g-2$ of the electron and muon (top left) and fixed-target experiments (light $m_{A_\td}$ and moderate $\epsilon$).  The light gray region shows limits from the BABAR $\Upsilon$(3S)$\rightarrow \gamma \mu^+ \mu^-$ search; the direct search limits are model-dependent and must be interpreted on a case-by-case basis.}
   \label{Fig: Dark Photon Limits}
\end{figure}

The dark photon communicates with the Standard Model through kinetic mixing and experimental bounds on these interactions arise from tests of QED. The most model-independent bound comes from the virtual exchange of the $A_\td$ between SM fields.
The best limits arise from the constraints on the magnetic dipole moments of the $\mu$ and $e$ \cite{Pospelov:2008zw}.   The constraints can be expressed as
\begin{eqnarray}
\epsilon^2 F\Big(\frac{m_e^2}{m_{A_\td}^2}  \Big) < 1.5\!\times\!\! 10^{-8}\quad
\epsilon^2 F\Big(\frac{m_\mu^2}{m_{A_\td}^2} \Big) < 6.4\!\times\!\! 10^{-6},
\end{eqnarray}
where 
\begin{eqnarray}
F(x)= \int_0^1\!\!dz \frac{ 2 z(1-z)^2}{(1-z)^2 + z/x} .
\end{eqnarray}
Fig.~\ref{Fig: Dark Photon Limits} shows that $g-2$ limits are most important at low $m_{A_\td}$ and large $\epsilon$.

For larger values of $m_{A_{\td}}$, constraints arise from precision electroweak interactions, which depend on the gauge terms of the Lagrangian.  The kinetically-mixed $U(1)_\td$ only alters the neutral currents and the Lagrangian for this sector is 
\begin{eqnarray}
\nonumber
\LL _{\text{Gauge}} &=& -\frac{1}{4}  \left( 
F^2_{\td \mu\nu} + B_{\mu\nu}^2 - 2 \epsilon F^{\mu\nu}_\td B_{\mu\nu} + W_{3 \mu\nu}^2 \right)\\
\nonumber
&&+ \frac{m_{Z^0}^2}{2} \left(1 + \frac{2h^0}{v}\right) Z^2  + \frac{m_{A_\td}^2}{2} \left(1 + \frac{\sqrt{2}\phi^0}{ f_\phi}\right) A^2_\td\\
&&+ A_\td J_\td + A_{\text{EM}} J_{\text{EM}} +  Z J_Z.
\label{eq: lagrangian}
\end{eqnarray}
The precision electroweak constraints have not been performed for dark photons with masses between $1 \GeV$ and $100 \GeV$.  In addition to oblique corrections, there is a non-oblique correction coming from the contribution of the dark photon to precision electromagnetic observables, such as differential Bhabha scattering.
A full analysis is beyond the scope of this paper and will be performed in \cite{upcoming2}.  
This article uses a bound on $\epsilon$ of \cite{Feldman:2007wj,Cassel:2009pu}
\begin{eqnarray}
\epsilon \lsim 1\times 10^{-2} .
\end{eqnarray}
This constraint becomes more important than the muon $g-2$ limit when $m_{A_{\td}} \gsim 250 \MeV$.

The width of the $Z^0$ is altered by the presence of the dark sector.  The interaction between the $Z^0$ and the dark current in the canonically normalized mass eigenstate basis is
\begin{equation}
\LL_{\text{Zd}} \simeq \Bigg(\epsilon s_{\theta} \frac{m^2_{A_\td}}{m_{Z^0}^2} \Bigg) Z^0_{\mu} J^{\mu}_\td,
\end{equation}
to lowest order in $\epsilon$ and $ m_{A_\td}^2/m_{Z^0}^2$.  The width of the $Z^0$ decay into the dark sector is 
\begin{eqnarray}
\label{Eq: Z Branching}
\Gamma(Z^0 \rightarrow \bar{\Psi}_L \Psi_L)  \simeq \frac{N_c g_\td^2 \epsilon^2 s_{\theta}^2 m_{A_\td}^4}{12 \pi m_{Z^0}^3} 
\simeq \frac{N_c s_{\theta}^2}{12 \pi} \frac{m_{A_\td}^8}{m_{Z^0}^3 f_{\text{eff}}^4}.
\end{eqnarray}
Any additional $Z^0$ decay mode cannot have a branching ratio of more than 0.18\% \cite{PDG}.   This sets a limit on the dark photon mass of 
\begin{eqnarray}
\epsilon\lsim \frac{0.045}{N_c^{\half}}\left(\frac{m_\pid}{70 \GeV}  \right) \left( \frac{100\GeV}{m_{A_\td} }\right)^3, 
\end{eqnarray}
which is never a constraint.

\subsection{Limits from Direct Production}
\label{Sec: DirectProd}

The allowed parameter space shown in Fig.~\ref{Fig: Dark Photon Limits} can be further constrained by searches for the direct production of the $A_\td$  \cite{ArkaniHamed:2008qp,Batell:2009yf,Essig:2009nc,Reece:2009un,Baumgart:2009tn}.  In this section, we outline the prospects for such searches and the challenges of translating the experimental bounds to theoretical constraints in composite models.   

If the dark photon is the lightest state in the dark sector ($m_{A_\td} < \Lambda_\td$), then it will decay directly to the SM.  Such a light $A_\td$ will be dominantly produced from $e^+e^- \rightarrow \gamma A_{\td}$ and from the decays of the dark hadrons.  Once produced, the dark photon will decay promptly; for example,
\begin{equation}
\Gamma(A_\td \rightarrow \ell^+ \ell^- ) \simeq \frac{1}{3} \epsilon^2 \alpha c_{\theta}^2 m_{A_\td} \simeq 30 \text{ eV}
\end{equation}
for the benchmark model CiDM$_1$.  Hadronic decays are also allowed, but are subdominant to the lepton decays, except near resonances \cite{Batell:2009yf}.

When the dark photon is heavier than $\Lambda_\td$, it can either decay to dark mesons or directly to SM leptons.  However, the coupling of the dark photon to the electromagnetic current is suppressed by a factor of $\epsilon$ relative to the coupling to the dark quarks
\begin{equation}
\LL_{\text{dd}} + \LL_{\text{dem}} \simeq A_{\td \mu} J_{\text{d}}^{\mu} + \epsilon c_{\theta} A_{\td \mu} J_{\text{em}}^{\mu}. 
\end{equation}
In this limit, the branching fraction into SM leptons is negligible:
\begin{equation}
\label{Eq: BranchingRatio}
\text{Br}(A_{\td} \rightarrow \ell\bar{\ell}) \simeq \frac{\epsilon^2 c_{\theta}^2 g^2}{N_c g_d^2}
\simeq \frac{64 g^2 c_{\theta}^2}{N_c} \frac{m_{\pid}^4}{f_{\text{eff}}^4}
\simeq 4.0\!\times\! 10^{-4}
\end{equation}
for $N_c =4$ and the benchmark model CiDM$_1$.  The dark photon preferentially decays to the dark quarks, which first parton shower, then hadronize, and finally cascade decay back to SM particles.  

The most common mesons formed in the hadronization process will be the lightest in the spectrum.  
There are no light pseudo Goldstone bosons in this theory because there is only one light quark.  Therefore, the lightest meson is $\eta'_\td$ ($0^{-+}$), in analogy with QCD.  Using the SM $\eta'$ as a prototype and the SM $a_0$ meson as a typical hadronic state, the mass of the $\eta'_\td$ is estimated to be 
\begin{eqnarray}
m_{\eta'_\td} \simeq\frac{\sqrt{3}}{\sqrt{N_c}} \frac{m_{\eta'}   }{m_{a_0}} \Lambda_\td \simeq \frac{ 1.7 \Lambda_\td}{\sqrt{N_c}}.
\end{eqnarray}
The $\eta'_\td$ becomes light in the large $N_c$ limit.  For $N_c \gtrsim 10$, the dark photon can decay to the $\eta'_\td$.
However, $\eta'_\td$ is cosmologically stable  because it has a chirality suppressed decay to electrons and primarily decays via a loop-induced process  to two leptons, which dominates over the four body decay $\eta'_\td \rightarrow A^*_\td A^*_\td$ \cite{Batell:2009yf,Schuster:2009au}.   The decay width is suppressed by an additional factor of $(\Lambda_\td/f_\phi)^2$ relative to the $\phi$ decay mode of  \cite{Batell:2009yf,Schuster:2009au,Schuster:2009fc} because the $\eta'_\td$ has to mix with the $\phi$ to mediate the decay.  The resulting decay width is
\begin{eqnarray}
\Gamma(\eta'_\td \rightarrow e^+ e^-)&\simeq&  \frac{\epsilon^4 \alpha^2 m_e^2 m_{\eta'_\td}  \Lambda_\td^2}{(4\pi)^3 f_\phi^4}
\ll \frac{1}{10^{10} \text{years}}.
\end{eqnarray}
The cosmological relic abundance of $\eta'_\td$ is sufficiently small to make up a small fraction of the matter density of the Universe. 

Unlike the $\eta'_\td$, the next-lightest meson, $\omega_\td$ ($1^{--}$), can have prompt decays.  The mass of the $\omega_\td$ also becomes small in the large $N_c$ limit.  The $\omega_\td$ will decay to SM leptons by mixing with the dark photon.  Approximating the mixing angle by
\begin{equation}
\theta_{\omega_\td} \simeq \frac{m_{\omega_\td}^4}{(m_{\omega_\td}^2+m_{A_\td}^2)^2},
\end{equation}
the decay width is  \cite{Schuster:2009au}
\begin{equation}
\Gamma(\omega_\td \rightarrow e^+ e^-\!) \simeq\! \frac{\epsilon^2\alpha g_{\td}^2  c_{\theta}^2 }{3} \frac{m_{\omega_\td}^5}{m_{A_\td}^4}
\simeq \frac{\alpha c_{\theta}^2}{3} \frac{\Lambda_\td^5}{f_{\text{eff}}^4}
\simeq \frac{1}{20 \text{ m}}
\end{equation}
for the CiDM$_1$ benchmark point in Table \ref{Tab: Benchmarks}.   Therefore, if the $\omega_\td$ is produced, it will decay to two leptons with a long displaced vertex.   

The dark photon will decay promptly to leptons  if $m_{A_\td} \ll \Lambda_\td$ and both the BABAR  \cite{BaBar:2009cp} and CLEO \cite{Love:2008hs} searches for $\Upsilon$(3S)$\rightarrow\gamma \mu^+\mu^-$ may be used to set bounds.  However, when the dark photon decay channels are closed, the muon decay channels are often closed as well because $\Lambda_\td \lsim 2 m_\mu$.  
 The estimated bounds from these searches are shown in Fig.~\ref{Fig: Dark Photon Limits}.     It may also be possible to use the $\Upsilon(1S)\rightarrow \gamma +X$, where $X$ is invisible, when the $A_\td$ decays outside the detector \cite{Rubin:2006gc} .
 
The best chance of discovering the dark sector is by directly producing the $A_\td$ at low-energy lepton colliders.  BABAR has recently searched for the $A_\td$ in the $4 \ell$ channel \cite{Aubert:2009pw}
 and future work is being pursued at a myriad of experiments \cite{Bjorken:2009mm, DarkForcesWorkshop}.  The searches are complicated because the decay of the $A_\td$ back to leptons is suppressed by the factor in Eq. \ref{Eq: BranchingRatio}.  To gain efficacy, it is necessary to use the decay into the dark quarks.    
For $m_{A_\td}/\Lambda_\td \lsim \OO(10)$, the number of dark hadrons is moderate.   Because the fluctuations in the number of hadrons is non-Gaussian, the cost of fragmenting to two $\omega_{\td}$ mesons is not limiting and it is possible to set limits using the BABAR and CLEO searches.  When  $m_{A_\td}/\Lambda_\td \gg 1$, there can be a large number of dark hadrons in the decay products of the $A_\td$, and an inclusive, multi-lepton search is necessary.  When the dark photon decays through the hadronic channel, the analysis becomes more challenging because it is necessary to know how the dark partons fragment into dark hadrons and then decay down to the Standard Model.  Setting limits on this model is beyond the scope of this work because of the significant uncertainties in the hadronic spectrum.  

In addition to low energy searches for the decay products for the $A_\td$, high energy colliders provide a useful laboratory.
LEP-I can search for rare decays down to  the $\Br(Z^0) \lsim 10^{-5}$.  From Eq. \ref{Eq: Z Branching}, this corresponds to 
masses of the $A_\td \gsim 4 \GeV$.  When the $A_\td$ decays with a mass $m_{A_\td}\gg \Lambda_\td$, it decays into a pair of dark quarks, $\Psi_L \bar{\Psi}_L$, and proceeds to shower and hadronize.    Future studies of LEP2 are needed to determine the relevant final states and the procedure necessary to set limits on the hetrogeneous final states. 

\section{Discussion}
\label{Sec: Discussion}

This article introduced a composite inelastic dark matter model with dominant inelastic scattering
off of nuclei and a subdominant elastic scattering component.  The subdominant elastic component is a signature of the symmetry structure of the model and is a critical feature to measure.  It was found that parity violating effects can be nearly maximal, with   
\begin{eqnarray}
\thpv \simeq \Theta_\td \frac{m_L}{\Lambda_\td} \lsim 0.08.
\end{eqnarray}
Discovering the elastic subcomponent will place a lower limit on $m_L$ and  could sharpen the Standard Model's strong CP problem and flavor structure.

Directional detection experiments, which measure both the energy and direction of recoiling nuclei in detectors \cite{Spergel:1987kx,Gondolo:2002np,Finkbeiner:2009ug}, can be used to distinguish the elastic and inelastic scattering components by looking for large-angle scattering events  \cite{Lisanti:2009vy}.  There is an upper bound on the allowed scattering angle, which depends on the types of interactions that are allowed.  In particular, 
\begin{equation}
\cos \gamma_{\text{max}} = \frac{v_{\text{esc}} -v_{\text{min}}}{v_e},
\end{equation}
where $\gamma$ is the angle between the direction of the Earth's velocity and the recoiling nucleus in the lab frame
and $v_{\text{min}}$ is the minimum velocity to scatter at a given recoil energy,
\begin{eqnarray}
v_\text{min}(E_R) = \begin{cases}
\sqrt{\frac{m_N E_R}{2 \mu^2}} & \text{elastic}\\
\frac{1}{\sqrt{2 m_N E_R}} \Big(\frac{m_N E_R}{\mu} + \delta m\Big) & \text{inelastic}
\end{cases} \,,
\end{eqnarray}
and $\mu$ is the reduced mass of the dark matter-nucleus system.  Inelastic scattering events have a much smaller $\cos \gamma_{\text{max}}$ than elastic scattering events and these two types of interactions can be distinguished with the next generation of directional detection experiments \cite{Ahlen:2009ev,Sciolla:2009fb, Miuchi:2007ga,Burgos:2008mv,Santos:2007ga}.

The best fit for CiDM had  $\delta m \sim 100 \keV$ and $m_\pid \sim 70 \GeV$.  This leads to an estimate of
\begin{eqnarray}
\Lambda_\td \simeq \sqrt{ \delta m \,  m_\pid} = 150 \MeV
\end{eqnarray}
for the dark sector confining scale.  The lower bound is $\Lambda_\td \gsim  70 \MeV/\sqrt{\kappa}$ and the upper bound is $\Lambda_\td \lsim 240 \MeV/\sqrt{\kappa}$ (for Fig. \ref{fig: masslambda}, a value of $\kappa= 0.25$ is used).
This indicates that the dark matter form factor may be important in shaping the higher energy bins
of DAMA.  Applying a dark matter form factor might therefore change the allowed parameter range.
For Coulombically-bound dark matter, the form factor can be found by Fourier transforming the hydrogenic wave functions to get
\begin{eqnarray}
F_{\text{DM}}(q^2) = \frac{1}{ 1 + q^2 r_{\text{DM}}^2},
\end{eqnarray}
where $r_{\text{DM}}^{-1}= \lambda_\td m_L$ is the Bohr radius of the Coulombically-bound state and $\lambda_\td$ is the 't Hooft coupling evaluated at the Bohr radius.   The strongly interacting form factors can be estimated by extrapolating $r_{\text{DM}}^{-1}\rightarrow \Lambda_\td$ and behave similarly to having $m_{A_\td} \sim r_{\text{DM}}^{-1}$.

CiDM models have sub-components to the dark matter that are not the $\pid$. There are roughly three classes of particles:
the $\rhod$, multiple heavy quark mesons ({\it e.g.}, $\Psi_H \Psi_H \bar{\Psi}_L \bar{\Psi}_L$ states), and baryons ({\it e.g.}, $\Psi_H \cdots \Psi_H$ states).    The relative populations of these states is determined by the interactions in the early Universe \cite{Alves:2009nf, Alves:2009XX}.  Detecting the latter two classes of particles would be a clear indication of composite inelastic dark matter.  The signature will be striking because the mass of the dark matter subcomponents would be near-integer multiples of the $\pid$ mass, ending at $N_c m_\pid$.  

The collider signatures of CiDM are challenging because many of the decay products have long lifetimes and give rise to extremely displaced vertices or missing energy.   To interpret the results from current $e^+e^-$ colliders, it is necessary to have better estimates for the dark hadron multiplicity distributions from dark photon decays.  If there are dynamics that stabilize the vev of both the Standard Model Higgs and the dark sector Higgs, then it is possible to produce dark sector states at the Tevatron and LHC in the decays of electroweak scale particles.  The phenomenology of these events will be similar to that of hidden valley theories, with the majority of parameter space giving rise to extremely displaced vertices \cite{Strassler:2006im,Strassler:2006ri,Han:2007ae}.  Beam dump experiments are ideally suited for identifying leptons from particles with a finite lifetime and provide the best prospect for discovering the dark sector through direct production.  Proposals for these experiments are presently underway \cite{DarkForcesWorkshop}. 

\section*{Acknowledgements}
We thank Philip Schuster for numerous illuminating discussions and thank Rouven Essig for providing the experimental limits on the $m_{A_\td}-\epsilon$ plane.
We also thank Spencer Chang, Graham Kribs, Tuhin Roy, Aaron Pierce, Neal Weiner, Matt Reece, Natalia Toro, and Liam Fitzpatrick for useful discussions.
ML  and JGW are supported by the US DOE under contract number DE-AC02-76SF00515 and receive partial support from the Stanford Institute for Theoretical Physics.  ML is supported by an NSF fellowship.
JGW is partially supported by the US DOE's Outstanding Junior Investigator Award.  


\end{document}